\documentclass[12pt,dvips]{article}

\usepackage{rotating}
\usepackage{axodraw} 
\usepackage{epsfig}  
\usepackage{epsf}    
\usepackage{psfig}   
\usepackage{cite}    

\newcommand{\imag}{\Im {\rm m}}
\newcommand{\real}{\Re {\rm e}}

\def\lsim{\stackrel{<}{{}_\sim}}

\begin{document}

\begin{flushright}
CERN-PH-TH/2005-112 \\
hep-ph/0507046 \\
July 2005
\end{flushright}

\begin{center}
{\bf {\LARGE Resonant CP Violation in Higgs Radiation }}\\[3mm]
{\bf {\LARGE at {\boldmath $e^+e^-$} Linear Collider}}
\end{center}

\bigskip

\begin{center}
{\large John Ellis$^{\,a}$, Jae Sik Lee$^{\,b}$
                                       and Apostolos Pilaftsis$^{\,b}$}
\end{center}

\begin{center}
{\em $^a$Theory Division, Physics Department, CERN, CH-1211 Geneva 23,
Switzerland}\\[2mm]
{\em $^b$School of Physics and Astronomy, University of Manchester}\\
{\em Manchester M13 9PL, United Kingdom}
\end{center}

\bigskip\bigskip\bigskip

\centerline{\bf ABSTRACT}
\noindent
We study  resonant CP violation in the  Higgsstrahlung process $e^+e^-
\rightarrow  H_{1,2,3}\  (Z  \to  e^+e^-,\mu^+\mu^-)$  and  subsequent
decays  $H_{1,2,3} \to  b \bar{b},\  \tau^-\tau^+$, in  the  MSSM with
Higgs-sector  CP violation  induced  by radiative  corrections.  At  a
high-energy $e^+e^-$  linear collider, the recoil-mass method enables
one  to determine the  invariant mass  of a  fermion pair  produced by
Higgs decays  with a precision as good as  1  GeV. Assuming an
integrated luminosity  of 100 fb$^{-1}$,  we show that  the production
lineshape of  a coupled system  of neutral Higgs bosons  decaying into
$b\bar{b}$ quarks  is sensitive to the  CP-violating parameters.  When
the Higgs bosons decay into  $\tau^-\tau^+$, two CP asymmetries can be
defined  using the  longitudinal and  transverse polarizations  of the
$\tau$  leptons.  Taking  into account  the constraints  from electric
dipole moments, we  find that these CP~asymmetries can  be as large as
80\%, in a tri-mixing scenario where all three neutral Higgs states of
the MSSM are nearly degenerate and mix significantly.

\newpage

\section{Introduction}
\label{sec:introduction}

A future $e^+e^-$ linear collider, such as the projected International
Linear Collider (ILC),  will have the potential to  probe the Standard
Model (SM)  Higgs sector with higher precision  than its predecessors,
the Tevatron collider at Fermilab  and the Large Hadron Collider (LHC)
at  CERN.   The present  direct  search  limits  from the  search  for
Higgsstrahlung~\cite{HS} at LEP show that the SM Higgs boson should be
heavier than about 114~GeV~\cite{LEPHWG}.   This lower limit is within
the   mass  range   favoured  indirectly   by   precision  electroweak
measurements~\cite{LEPEWWG}.  Future refinements  of these  direct and
indirect limits  will be very  crucial for identifying  the underlying
structure of  the fundamental  Higgs sector, within  either the  SM or
some non-minimal model of Higgs physics.

One well-motivated model of physics beyond the SM is the minimal
supersymmetric extension of the Standard Model (MSSM)~\cite{HPN}.  The
MSSM predicts three neutral Higgs states.  In the presence of explicit
CP-violating sources, such as complex soft squark masses, gaugino 
masses
and trilinear couplings, all the three neutral Higgs bosons mix through
CP-violating quantum effects~\cite{APLB,PW,Demir,CDL,CEPW} to mass
eigenstates, $H_{1,2,3}$, of indefinite CP. In this CP-violating MSSM, 
one
interesting possibility is that the lightest neutral Higgs boson could 
be
considerably lighter than 114~GeV~\cite{CPX,CEMPW}, at moderate values 
of
$\tan\beta \stackrel{<}{{}_\sim} 10$, for relatively light charged
Higgs-boson masses, $M_{H^\pm} \sim 130$--170~GeV. Alternatively, if
$\tan\beta \stackrel{>}{{}_\sim} 40$, all three neutral Higgs states
$H_{1,2,3}$ may have similar masses and strongly mix with each other
dynamically, through CP-violating off-diagonal absorptive self-energy
effects~\cite{APNPB,ELP1}. Such a scenario was studied
in~\cite{ELP1,ELP2,ELP3} and termed the CP-violating tri-mixing 
scenario
of the MSSM.

The general unconstrained  MSSM has dozens  of additional CP-violating
phases  beyond the Kobayashi--Maskawa phase   of the SM. These  phases
appear in  complex soft SUSY-breaking  mass  parameters of sfermions 
and
gauginos and in complex trilinear Yukawa couplings. They have a wealth
of phenomenological  implications.  In particular, they  give  rise to
signatures of CP violation    in  sparticle production and    decay at
high-energy colliders~\cite{CPdirect,CPsoft1,CPsoft2}, have observable
effects  on electric  dipole  moments  (EDMs)~\cite{EDM1,EDM2,CKP} and
$B$-meson   decays~\cite{Bmeson1,DP}, and  might  constitute the extra
ingredients needed for electroweak baryogenesis~\cite{baryog}.

In this   paper  we  study resonant  CP-violating  phenomena    in the
Higgsstrahlung mechanism for producing neutral MSSM  Higgs bosons at a
high-energy $e^+e^-$ linear collider.  Although our study is performed
within    the   radiative       CP-violating   framework    of     the
MSSM~\cite{APLB,PW,Demir,CDL,CEPW,INhiggs,KW,HeinCP,CEPW2},        the
results of  our analysis  would   also be  applicable to  more general
CP-violating two-Higgs-doublet  models~\cite{Maria} in which all three
neutral Higgs bosons mix strongly.  This work extends previous studies
of   the masses, couplings,  production   and  decays of the  mixed-CP
neutral   Higgs bosons  $H_{1,2,3}$,  with  a  view   to searches   at
LEP~\cite{CPX},                                                    the
LHC~\cite{CPX,CEMPW,CHL,CPpp,CFLMP,Akeroyd,KMR,ELP3},              the
ILC~\cite{CPee}, a $\mu^+\mu^-$  collider~\cite{CPmumu} and  a Photon 
Linear
Collider (PLC)~\cite{CPphoton,PPTT,GKS,CKLZ,ELP2}.  As in      our
previous  works~\cite{ELP1,ELP3,ELP2}, we present a complete treatment
of loop-induced    CP   violation and   Higgs  tri-mixing,   including
off-diagonal absorptive effects in the resummed Higgs-boson propagator
matrix~\cite{APNPB}.
Complementary to the previous studies at the ILC~\cite{CPee}, our focus 
here
is on analyzing the production lineshape of a coupled system of
neutral Higgs bosons in the Higgsstrahlung process, as well as the
construction of feasible CP asymmetries which could be probed
experimentally.

Higgs-boson     production   via   the   Higgsstrahlung        process
$e^+e^-\rightarrow    H_i\,Z$~\cite{HS}, where the $Z$ boson decays 
into
electron  or muon pairs, offers   a unique environment for determining
the masses and widths of  the neutral Higgs bosons  by the recoil-mass
method~\cite{GAbia,BBDS}.  Thanks to the excellent energy and momentum
resolution of electrons and muons coming from the $Z$-boson decay, the
recoil  mass   against   the  $Z$  boson,   $p^2=s-2\cdot\sqrt{s}\cdot
E_Z+M_Z^2$, can be reconstructed with a precision as good as 1
GeV.  Here $s$ and  $E_Z$ are the  the collider centre-of-mass  energy
squared and the energy of the $Z$ boson, respectively.
 
As  mentioned above,  our  focus  is  on the CP-violating   tri-mixing
scenario of the MSSM,  in which  all  three neutral Higgs  bosons have
similar masses and mix  strongly with each  other.  In  particular, we
examine the  production  lineshape of  the coupled   system of neutral
Higgs bosons  in   this  tri-mixing   scenario.  The  lineshape    for
$H_{1,2,3} \to  {\bar  b}b$ decays is  sensitive   to the CP-violating
parameters of the model.  Moreover,   employing the longitudinal   and
transverse polarizations of the tau leptons  coming from the decays of
Higgs bosons, we can measure  CP asymmetries which can  be as large as
80\%, without violating EDM constraints.

The layout of the   paper is as follows.  In  Section~2, we study  the
Higgsstrahlung   process at an   $e^+e^-$ collider  and the subsequent
Higgs-boson  decays,   $H_i\rightarrow  f\bar{f}$   with   $f=b$   and
$\tau^-$. We  define the individual cross sections that depend on
the   longitudinal  and transverse polarizations of the final fermions.
In Section~3, we   consider the constraints from the
non-observation  of an  EDM in  the  Thallium atom on  the relevant CP
phases in the tri-mixing scenario.   
In Section~4, we construct two CP asymmetries and present
numerical examples for the two final  states $f\bar{f} = b{\bar b}$ and
$\tau^- \tau^+$.  Our conclusions are given in Section~5.

\section{The Process {\boldmath $e^+e^-\rightarrow H_i\,Z\rightarrow 
[f(\sigma)\bar{f}(\bar{\sigma})]\,Z$}   }
\label{sec:higgsstrahlung}

\begin{figure}[t]
\vspace*{1.5cm}
\begin{center}
\begin{picture}(300,100)(0,0)

\Text(-40,70)[]{$s\rightarrow$}

\ArrowLine(0,100)(50,70)\ArrowLine(10,100)(40,82)
\Text(10,120)[]{$e^-(p_1,\omega)$}

\ArrowLine(50,70)(0,40)\ArrowLine(10,40)(40,58)
\Text(10,30)[]{$e^+(p_2,\bar\omega)$}

\Photon(50,70)(120,70){4}{6}
\Text(85,85)[]{$Z^*(\sqrt{s})$}


\Photon(120,70)(200,130){4}{8}\ArrowLine(140,100)(180,128)\Text(155,120)[]{$k$}
\Text(225,130)[]{$Z(k,\lambda)$}

\DashLine(120,70)(200,10){3}\GOval(160,40)(10,10)(360){0.9}
\Text(145,65)[]{$H_i$}\Text(185,35)[]{$H_j$}
\ArrowLine(132,42)(172,12)\Text(145,20)[]{$p$}

\ArrowLine(200,10)(250,40)
\ArrowLine(210,22)(240,40)\Text(220,40)[]{$l_1$}
\Text(270,40)[]{$f(l_1,\sigma)$}

\ArrowLine(250,-20)(200,10)
\ArrowLine(210,-5)(240,-23)\Text(220,-20)[]{$l_2$}
\Text(270,-20)[]{$\bar{f}(l_2,\bar\sigma)$}

\GOval(200,10)(5,5)(360){0.9}

\end{picture} \\
\end{center}
\medskip
\noindent
\caption{\it The dominant mechanism contributing to
the process $e^+e^-\rightarrow H_i\,Z\rightarrow 
[f(\sigma)\bar{f}(\bar{\sigma})]\,Z$.}
\label{fig:eezff}
\end{figure}
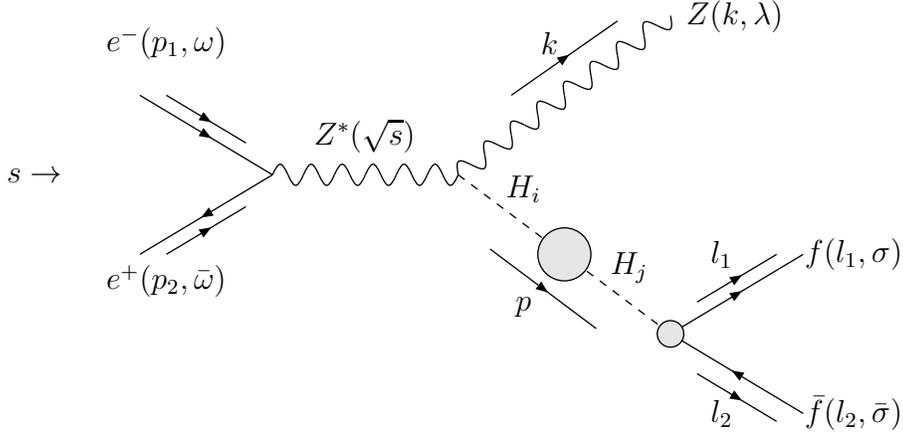

The   general       helicity     amplitude    for       the    process
$e^-(p_1,\omega)\,e^+(p_2,\bar\omega)\rightarrow  H_i(p)\,Z(k,\lambda)
\rightarrow   [f(l_1,\sigma)\bar{f}(l_2,\bar{\sigma})]\,Z(k,\lambda)$,
depicted in Fig.~\ref{fig:eezff}, may be written as
\begin{equation}
{\cal M}_\omega^{\lambda\sigma}\ =\ \frac{-g^2g_f
M_W}{c_W^3\sqrt{s}\sqrt{p^2}}D_Z(s)
\langle\omega;\lambda\rangle_\Theta \,
\left(\sum_{i,j}\langle\sigma\rangle^f_{ij}\right)\,
\delta_{\omega-\bar\omega}\delta_{\sigma\bar\sigma}.
\label{eq:helamp}
\end{equation}
The  four-momenta and helicities of  the initial electron and positron
are denoted  by $(p_1,\omega)$ and  $(p_2,\bar\omega)$,  respectively,
and  $s \equiv  (p_1+p_2)^2$.  We  denote  the helicities   of $f$ and
$\bar{f}$ by $\sigma$ and $\bar\sigma$,  and that of  the $Z$ boson by
$\lambda$.   Also, $\sigma=+(-)$  stands  for   a right- (left)-handed
particle and $\lambda=\pm$  and $\lambda=0$ for the  transverse (right
and left helicities) and longitudinal polarizations, respectively. The
four-momentum  of the intermediate Higgs boson  is  denoted by $p$ and
those  of     the   final    fermions by    $l_1$   and    $l_2$  with
$p=l_1+l_2$. Finally, $\Theta$  is the  angle between  $\bf{p_1}$  and
$\bf{k}$ where the four-momentum of the $Z$ boson is $k=(E_Z,\bf{k})$.

In~(\ref{eq:helamp}), $D_Z(s)$   is  the $s$-normalized  Breit--Wigner
propagator for the $Z$ boson:
\begin{equation}
D_Z(s)\ =\  \frac{s}{s-M_Z^2+iM_Z\Gamma_Z},
\end{equation}
and              $\langle\omega;\lambda\rangle_\Theta$             and
$\langle\sigma\rangle^f_{ij}$ describe the reduced amplitudes
\begin{eqnarray}
\langle\omega;\lambda=0\rangle_\Theta &=& -(v_e+\omega a_e)\,
\frac{\omega E_Z}{M_Z} s_\Theta\,, \qquad
\langle\omega;\lambda=\pm\rangle_\Theta\ \, =\ \, (v_e+\omega a_e)\,
\frac{1+\omega\lambda c_\Theta}{\sqrt{2}}\,,
\nonumber \\
\langle\sigma\rangle^f_{ij} &=&
g_{H_iVV}D_{ij}(p^2)\, (\sigma\beta 
g^S_{H_j\bar{f}f}-ig^P_{H_j\bar{f}f})\;,
\end{eqnarray}
where  $s_\Theta\equiv\sin\Theta$,   $c_\Theta\equiv\cos\Theta$    and
$v_e=-1/4+s_W^2$  and $a_e=1/4$.  This  result  is consistent with the
one given  in~\cite{HagiwaraStong,HIKK}.   For the definitions of  the
couplings,  the  threshold  corrections that  are  enhanced for  large
values of  $\tan\beta$  for $f=b\,,\tau$,   and the full   $3\times 3$
propagator matrix $D_{ij}(p^2)$, we refer to~\cite{ELP1,CPsuperH}.

The differential cross-section, after  integrating over $c_\Theta$, is
given by
\begin{eqnarray}
p^2\frac{d\sigma}{dp^2}&=&
{\sigma_{HZ}^{\rm SM}(p^2)}\times
\frac{N_f\,g_f^2 \beta_f(p^2)}{16\pi^2}
\left\{
(1+{P_L\bar{P_L}})C^f_1(p^2)+
{(P_L+\bar{P_L})}{C^f_2(p^2)} \right.\nonumber \\
&&+\left.
{P_T\bar{P_T}}\left[\cos(\alpha-\bar\alpha)C^f_3(p^2)
+\sin(\alpha-\bar\alpha){C^f_4(p^2)}\right] \right\}\,,
\label{eq:cx}
\end{eqnarray}
where $N_f$ is the color factor of  the final-state fermion: $N_f = 3$
for the  $b$ quark and 1 for  the $\tau$  lepton, $\beta^f(p^2) \equiv
\sqrt{1-4m_f^2/p^2}$. Moreover,  $\sigma_{HZ}^{\rm SM}(p^2)$ is the SM
cross section for the Higgsstrahlung of an  off-shell Higgs boson with
mass $\sqrt{p^2}$, i.e.,
\begin{equation}
{\sigma_{HZ}^{\rm SM}(p^2)}\
=\ \frac{g^4(v_e^2+a_e^2)}{192\pi c_W^4 s}\,
\frac{\lambda^{1/2}(p^2)[\lambda(p^2)+12M_Z^2/s]}
{(1-M_Z^2/s)^2+M_Z^2\Gamma_Z^2/s^2}\ ,
\end{equation}
where   $\lambda   (p^2) =   (1-p^2/s-M_Z^2/s)^2-4p^2M_Z^2/s^2$  is  a
kinematic phase-space function.
Then, with the definition
\begin{equation}
\langle\sigma\rangle^f\ \equiv\ \sum_{i,j} 
\langle\sigma\rangle^f_{ij}\;,
\end{equation}
the polarization coefficients $C_i^f$ may conveniently be expressed as
follows:
\begin{eqnarray}
C^f_1(p^2) \!&=&\! \frac{1}{4}\left(\left|\langle + \rangle^f\right|^2
+\left|\langle - \rangle^f\right|^2\right)\,, \qquad\quad
C^f_2(p^2)\ =\ \frac{1}{4}\left(\left|\langle + \rangle^f\right|^2
-\left|\langle - \rangle^f\right|^2\right)\,, \nonumber \\
C^f_3(p^2) \!&=&\! -\frac{1}{2}\real\left(\langle + \rangle^f
\langle - \rangle^{f*}\right)\; ,  \qquad\qquad
C^f_4(p^2)\ =\ \frac{1}{2}\imag\left(\langle + \rangle^f
\langle - \rangle^{f*}\right)\; .
\end{eqnarray}

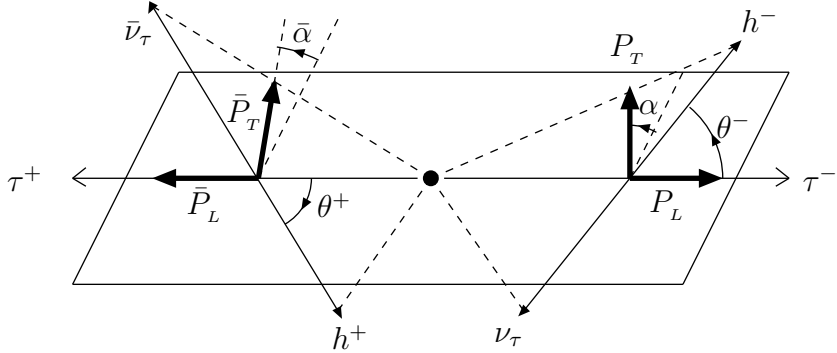
\begin{figure}[t]
\vspace*{1cm}
\begin{center}
\begin{picture}(300,100)(0,0)

\Line(15,10)(55,90)
\Line(55,90)(285,90)
\Line(285,90)(245,10)
\Line(245,10)(15,10)

\Text(-3,50)[]{$\tau^+$}
\Line(15,50)(145,50)
\Line(15,50)(20,54)
\Line(15,50)(20,47)

\Text(298,50)[]{$\tau^-$}
\Line(155,50)(285,50)
\Line(285,50)(280,54)
\Line(285,50)(280,47)

\Vertex(150,50){3}

\DashLine(90,80)(95,110){3}
\DashLine(85,50)(115,110){3}
\ArrowArc(85,50)(50,63,82)
\Text(102,105)[]{$\bar{\alpha}$}

\DashLine(225,50)(245,90){3}
\ArrowArc(225,50)(20,60,90)
\Text(233,77)[]{$\alpha$}

\SetWidth{2}
\Line(225,50)(225,80)\ArrowLine(225,80)(225,81)
\Text(225,99)[]{$P_{_T}$}

\Line(225,50)(255,50)\ArrowLine(255,50)(256,50)
\Text(240,40)[]{${P}_{_L}$}

\Line(85,50)(90,80)\ArrowLine(90,80)(91,86)
\Text(80,75)[]{$\bar{P}_{_T}$}

\Line(85,50)(50,50)\ArrowLine(50,50)(49,50)
\Text(65,40)[]{$\bar{P}_{_L}$}

\SetWidth{0.5}
\Line(225,50)(265,100)\ArrowLine(265,100)(266,101)\Text(275,110)[]{$h^-$}
\ArrowArc(225,50)(35,0,50)\Text(265,70)[]{$\theta^-$}
\DashLine(150,50)(265,100){3}
\Line(225,50)(185,0)\ArrowLine(185,0)(184,-1)\Text(180,-10)[]{$\nu_\tau$}
\DashLine(150,50)(185,0){3}

\DashLine(150,50)(45,115){3}
\DashLine(150,50)(115,0){3}
\Line(45,115)(115,0)\ArrowLine(45,115)(44,113)\ArrowLine(115,0)(116,-2)
\Text(40,105)[]{$\bar{\nu}_\tau$}
\Text(120,-10)[]{$h^+$}
\ArrowArcn(85,50)(20,0,300)\Text(114,40)[]{$\theta^+$}

\end{picture}\\
\end{center}
\medskip
\noindent
\caption{\it The $\tau^+\tau^-$ production plane in the Higgs-boson
rest frame, in the case when the $\tau$ leptons decay into hadrons
$h^\pm$ and neutrinos.  The longitudinal-polarization vector
$P_L(\bar{P}_L)$ and the transverse-polarization vector
$P_T(\bar{P}_T)$ with the azimuthal angle $\alpha(\bar\alpha)$ of
$\tau^-\,(\tau^+)$ are shown. }
\label{fig:tautau}
\end{figure}

In~(\ref{eq:cx}),   $P_L$ and   $\bar{P}_L$   are   the   longitudinal
polarizations  of the   final fermion $f$  and  antifermion $\bar{f}$,
respectively,   whereas  $P_T$ and    $\bar{P}_T$  are the degrees  of
transverse  polarization, with $\alpha$  and  $\bar{\alpha}$ being the
azimuthal angles with  respect to the $f$-$\bar{f}$  production plane.
We   depict  in  Fig.~\ref{fig:tautau} the   production  plane  in the
Higgs-boson  rest frame  in  the case when  $f=\tau^-$ and  the $\tau$
leptons    decay  into  charged     hadrons   $h^\pm$ and   neutrinos,
$\tau^\mp\rightarrow   h^\mp    \nu_\tau    (\bar{\nu}_\tau)$     with
$h^\pm=\pi^\pm$, $\rho^\pm$, {\it etc.}

Identifying the polarization analyser for $\tau^\mp$ as
\begin{equation}
\hat{a}^\mp=\pm\hat{h}^{\mp}\,,
\end{equation}
where $\hat{h}^\mp$ denote unit vectors parallel to the
$h^\mp$ momenta in the $\tau^\mp$ rest frame, we have
\begin{equation}
P_L=\cos\theta^-\,, \
P_T=\sin\theta^-\,, \
\alpha=\varphi^-\,; \ \
\bar{P}_L=\cos\theta^+\,, \
\bar{P}_T=\sin\theta^+\,, \
\bar{\alpha}=\varphi^+-\pi\,,
\end{equation}
where  $\theta^\pm$ and   $\varphi^\pm$ are  the polar  and  azimuthal
angles of  $h^\pm$, respectively, in the  $\tau^\pm$ rest frame.  With
this   identification,     the expression  in    the    curly brackets
of~(\ref{eq:cx}) becomes
\begin{eqnarray}
&&\hspace{-1cm}
(1+P_L\bar{P_L})C^f_1(p^2)+
(P_L+\bar{P_L})C^f_2(p^2)+
P_T\bar{P_T}\left[\cos(\alpha-\bar\alpha)C^f_3(p^2)
+\sin(\alpha-\bar\alpha)C^f_4(p^2)\right] \nonumber \\
&=&
(1+\cos\theta^-\cos\theta^+)C^f_1(p^2)+
(\cos\theta^-+\cos\theta^+)C^f_2(p^2) \nonumber \\
&&- \sin\theta^+\sin\theta^-\left[\cos(\varphi^--\varphi^+)C^f_3(p^2)
+\sin(\varphi^--\varphi^+)C^f_4(p^2)\right]\,.
\label{eq:taudecay}
\end{eqnarray}
We  observe that the  polarization coefficients $C_i^\tau(p^2)$ can be
determined  by  examining the   angular distributions of   the charged
hadrons coming  from  the $\tau$-lepton  decays;  $\tau^\mp\rightarrow
h^\mp \nu(\bar{\nu})$~\cite{TAUPOL}\footnote{ In  Refs.~\cite{TAUPOL},
the CP- and CP$\tilde{\rm T}$-odd  $C_2^f(p^2)$ coefficient is missing
since only one Higgs state was considered.}.

Finally,    for         our  phenomenological       discussion      in
Section~\ref{sec:examples},  it  proves more convenient to  define the
individual cross sections:
\begin{equation}
\hat{\sigma}^f_i(p^2)\ \equiv\
\sigma_{HZ}(p^2)\: \frac{N_fg_f^2\beta_f(p^2)}{16\pi^2}\: C_i^f(p^2)\,.
\label{eq:s14}
\end{equation}

\section{The Tri-mixing Scenario and the Thallium EDM}
\label{sec:thallium}

We take  the   following  parameter  set  for  numerical  examples  in
Section~\ref{sec:examples}:
\begin{eqnarray}
&&\tan\beta=50, \ \ M_{H^\pm}^{\rm pole}=155~~{\rm GeV},
\nonumber \\
&&M_{\tilde{Q}_3} = M_{\tilde{U}_3} = M_{\tilde{D}_3} =
M_{\tilde{L}_3} = M_{\tilde{E}_3} = M_{\rm SUSY} = 0.5 ~~{\rm TeV},
\nonumber \\
&& |\mu|=0.5 ~~{\rm TeV}, \ \
|A_{t,b,\tau}|=1 ~~{\rm TeV},   \ \
|M_2|=|M_1|=0.3~~{\rm TeV}, \ \ |M_3|=1 ~~{\rm TeV},
\nonumber \\
&&
\Phi_\mu = 0^\circ, \ \
\Phi_1 = \Phi_2 = 0^\circ\,.
\label{eq:tri}
\end{eqnarray}
We refer to this  scenario as the  tri-mixing scenario, since the mass
differences between three  neutral Higgs bosons  are comparable to the
decay widths and all  the three Higgs  bosons mix significantly in the
presence of   non-vanishing CP phases.  In  this  scenario, the common
third   generation  phase  $\Phi_A  =   \Phi_{A_t}    =  \Phi_{A_b}  =
\Phi_{A_\tau}$ and that of the gluino mass parameter $\Phi_3$ are free
parameters, which  are constrained by the  non-observation  of EDMs of
atoms, molecules, and  the neutron~\cite{CKP}.   The contributions  of
the   first  and second  generation  phases,  e.g.~$\Phi_{A_{e,\mu}}$,
$\Phi_{A_{d,s}}$ etc.,  to EDMs can be  drastically reduced either by 
making these phases  sufficiently   small, or if the first- and 
second-generation squarks and  sleptons are sufficiently heavy.  The 
impact   of these phases on the Higgs sector is negligible.

The EDM of the Thallium atom currently provides the best constraint on
the MSSM scenarios of our interest.  The  atomic EDM of $^{205}$Tl get
main contributions from two terms \cite{KL,PR}:
\begin{equation}
d_{\rm Tl}\,[e\,cm]\ =\ -585\cdot d_e\,[e\,cm]\:
-\: 8.5\times 10^{-19}\,[e\,cm]\cdot (C_S\,{\rm TeV}^2)+ \cdots\,,
\end{equation}
where $d_e$ denotes the electron EDM and $C_S$ is the coefficient of a
CP-odd      electron-nucleon     interaction    ${\cal 
L}_{C_S}=C_S\,\bar{e}i\gamma_5
e\,\bar{N}N$~\footnote{Our sign convention  for $C_S$  follows the one
given in~\cite{PR,AR}.}.   The  dots denote sub-dominant contributions
from  6-dimensional  tensor and   higher-dimensional  operators.   The
experimental $2-\sigma$ bound on the Thallium EDM is~\cite{THEDMEXP}:
\begin{equation}
|d_{\rm Tl}|\ \lsim\ 1.3\times 10^{-24}\,[e\,cm]\; .
\end{equation}

In the {\tt CPsuperH} \cite{CPsuperH} conventions and notations, 
the coefficient $C_S$ is given by
\begin{equation}
C_S\ =\ -(0.1\,{\rm GeV})\,\tan\beta\frac{m_e\pi\alpha_{\rm
em}}{s_W^2M_W^2}\sum_{i=1}^3
\frac{g_{H_igg}O_{ai}}{M_{H_i}^2}\ ,
\label{eq:cs}
\end{equation}
where
\begin{equation}
g_{H_igg}=\sum_{q=t,b}\left\{\frac{2}{3}g_{H_iqq}^S-\frac{v^2}{12}
\sum_{j=1,2} \frac{g_{H_i\tilde{q}_j^*\tilde{q}_j}}{m_{\tilde{q}_j}^2}
\right\}\ .
\end{equation}
The  Higgs-boson two-loop contributions to    the electron EDM   $d_e$
are~\cite{APEDM}:
\begin{eqnarray}
\left(\frac{d_e}{e}\right)^{\tilde{q}}&=&
\frac{3\alpha_{\rm
em}\,Q_q^2\,m_e}{32\pi^3}\sum_{i=1}^3\frac{g^P_{H_ie^+e^-}}{M_{H_i}^2}
\sum_{j=1,2} g_{H_i\tilde{q}_j^*\tilde{q}_j}\,F(\tau_{\tilde{q}_ji})\,,
\nonumber \\
\left(\frac{d_e}{e}\right)^q&=&
-\frac{3\alpha_{\rm em}^2\,Q_q^2\,m_e}{8\pi^2s_W^2M_W^2}
\sum_{i=1}^3\left[
g^P_{H_ie^+e^-} g^S_{H_i\bar{q}q}\,f(\tau_{qi})
+g^S_{H_ie^+e^-} g^P_{H_i\bar{q}q}\,g(\tau_{qi})
\right]\,,  \\
\left(\frac{d_e}{e}\right)^{\chi^\pm}&=&
-\frac{\alpha_{\rm em}^2\,m_e}{4\sqrt{2}\pi^2s_W^2M_W} \nonumber\\
&&\times\,
\sum_{i=1}^3\sum_{j=1,2}\frac{1}{m_{\chi^\pm_j}}\left[
g^P_{H_ie^+e^-} g^S_{H_i\chi^+_j\chi^-_j}\,f(\tau_{\chi_j^\pm i})
+g^S_{H_ie^+e^-} g^P_{H_i\chi^+_j\chi^-_j}\,g(\tau_{\chi_j^\pm i})
\right]\,,\nonumber
\label{eq:dee}
\end{eqnarray}
with $q=t,b$ and $\tau_{xi}=m_x^2/M_{H_i}^2$.
The total Higgs-mediated two-loop $(d_e/e)^H$ is given by the sum
\begin{equation}
  \label{totalEDM}
\left(\frac{d_e}{e}\right)^H\ =\
\left(\frac{d_e}{e}\right)^{\tilde{t}}+
\left(\frac{d_e}{e}\right)^{\tilde{b}}+
\left(\frac{d_e}{e}\right)^t+
\left(\frac{d_e}{e}\right)^b+
\left(\frac{d_e}{e}\right)^{\chi^\pm}\ +\ \dots\,,
\end{equation}
and the two-loop functions $F(\tau )$,  $f(\tau )$, and $g(\tau )$ may
be found in Ref.~\cite{APEDM}. The ellipses in~(\ref{totalEDM}) denote
other sub-dominant two-loop   contributions to EDM  that involve 
charged
Higgs and Higgsino effects~\cite{APEDMrest}.
 
We  note that  $C_S$ in~(\ref{eq:cs}) and  $(d_e/e)^H$ 
in~(\ref{eq:dee})
are  calculated at the electroweak  (EW)  scale, where the responsible
effective interactions are generated.
In order to calculate the running from the EW scale to the appropriate
low-energy    scale, the   anomalous    dimension (matrix)  should  be
considered.   We neglect this effect    by observing  that it can   be
absorbed in the evaluation of the matrix element
\begin{equation}
\langle N|\frac{\alpha_S}{8\pi}G^{a\,,\mu\nu}G^a_{\mu\nu} | N\rangle\
=\ -(0.1)\,{\rm GeV}\, \bar{N}N\,,
\end{equation}
which gives~(\ref{eq:cs}). We use $\alpha_{\rm em}=1/137$.

\begin{figure}[t]
\vspace{-0.5cm}
\centerline{\epsfig{figure=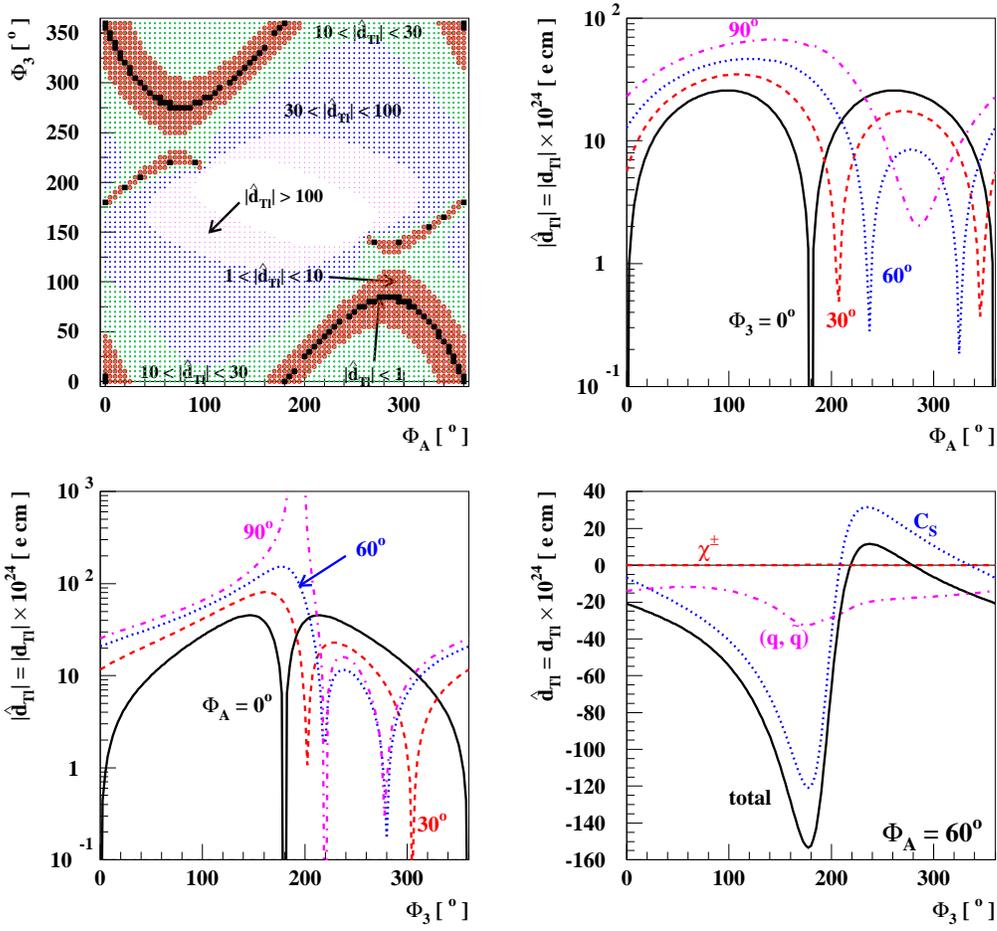,height=14cm,width=14cm}}
\caption{{\it 
The Thallium EDM $\hat{d}_{\rm Tl} \equiv d_{\rm Tl} \times 
10^{24}$~$e\,cm$ in the tri-mixing scenario. The
upper-left frame displays $|\hat{d}_{\rm Tl}|$ in the $(\Phi_A, 
\Phi_3)$ 
plane. The unshaded region around the point $\Phi_3=\Phi_A=180^o$ is
not theoretically allowed. 
The different shaded regions correspond to different ranges of 
$|\hat{d}_{\rm Tl}|$, as shown: specifically, $|\hat{d}_{\rm Tl}|<1$
in the narrow region denoted by filled black squares.
In the upper-right frame, we show $|\hat{d}_{\rm Tl}|$ as a function of
$\Phi_A$ for several values of $\Phi_3$.
In the lower-left frame, we show $|\hat{d}_{\rm Tl}|$ as a function of
$\Phi_3$ for four values of $\Phi_A$.
In the lower-right frame, we show the $C_S$ (dotted line) and
$(d_e/e)^{q,\tilde{q}}$ (dash-dotted line) contributions 
to $\hat{d}_{\rm Tl}$ separately
as functions of $\Phi_3$ when $\Phi_A=60^o$. As shown by the dashed 
line, the chargino contribution is negligible.
}}
\label{fig:p3pa}
\end{figure}

In   Fig.~\ref{fig:p3pa},  we  show    the   rescaled  Thallium    EDM
$\hat{d}_{\rm   Tl} \equiv d_{\rm   Tl} \times  10^{24}$  in  units of
$e\,cm$ in the $\Phi_A$-$\Phi_3$ plane (upper left)  and as a function
of $\Phi_A$  ($\Phi_3$) for several  values  of $\Phi_3$ ($\Phi_A$) in
the upper-right (lower-left)   frame. The lower-right frame shows  the
individual     contributions  as     functions    of    $\Phi_3$  when
$\Phi_A=60^\circ$.  We consider contributions  from the Higgs-mediated
two-loop $(d_e/e)^H$ and $C_S$, not  including other contributions. In
the  upper-left  frame, different ranges  of  $|\hat{d}_{\rm Tl}|$ are
shown explicitly by different shadings.
The blank unshaded  region around  the point $\Phi_3=\Phi_A=180^o$  is
not theoretically allowed since there large threshold corrections to 
the
bottom-quark Yukawa  coupling $h_b$ result  in a tachyonic  sbottom, a
complex or negative Higgs mass, a non-perturbative value of the Yukawa
coupling   $|h_b|>2$,  and/or a  failure  of the  iteration  method of
calculating the corrections.
We note that, in the region  $|\hat{d}_{\rm Tl}|<10$, the Thallium EDM
constraint can be evaded by assuming cancellations of less than 1 part
in 10 between the two-loop contributions considered here and possible
one-loop contributions not discussed here. As
mentioned above, such cancellations are always possible in a  general
unconstrained MSSM scenario,   where one-loop EDM  effects  depend  on
different CP-odd  phases related  to  the first and second generations  
of
squarks and sleptons. In our case, cancellations between the 
contributions
from $(d_e/e)^H$ and  $C_S$ are shown  in Fig.~\ref{fig:p3pa} as  dips
for specific  values of $\Phi_A$  and $\Phi_3$ in  the upper-right and
lower-left frames. These are responsible for  the narrow region filled
with black  squares   in the  upper-left  frame in which 
$|\hat{d}_{\rm Tl}| < 1\,e\,cm$. 
For  example,  the
lower-right frame  clearly   shows    the cancellation  between    the
contributions from $(d_e/e)^{q,\tilde{q}}$ and $C_S$ when $\Phi_3 \sim
220^\circ$ and $280^\circ$ for $\Phi_A=60^\circ$,  see also the dotted
line in the lower-left frame.

In   the  CP-violating   tri-mixing   scenarios  of   the  MSSM,   the
flavour-changing  neutral  current   couplings  of  the  Higgs  bosons
$H_{1,2,3}$  to down-type  quarks are  considerably enhanced  at large
values   of   $\tan\beta$~\cite{FCNC,Bmeson1},   i.e.~for   $\tan\beta
\stackrel{>}{{}_\sim} 40$.  These $\tan\beta$-enhanced Higgs couplings
can give  rise to potentially important constraints  on the parameters
of the CP-violating MSSM, which  arise from the non-observation of the
Higgs-mediated   $B$-meson   decay  $B_{s,d}   \to   \mu\mu$  at   the
Tevatron~\cite{TEVATRONmumu}.     However,   according   to    a   
detailed
study~\cite{DP},  the  derived  constraints  are highly  dependent  on
detailed aspects  of flavour physics, and may  be relaxed dramatically
for certain  choices of the soft  supersymmetry-breaking mass spectrum
that cause cancellations in the  unitarity sum over quark flavours. In
view of these and other theoretical uncertainties~\cite{EOS}, $B_{s,d}
\to \mu\mu$  decays do not  yet impose significant constraints  on the
parameter space relevant to this study.

\section{Numerical Examples}
\label{sec:examples}

For numerical studies, we take $\sqrt{s}=300$ GeV and consider four 
different
combinations of CP phases of $(\Phi_3,\Phi_A)$ in the tri-mixing 
scenario
(\ref{eq:tri}) that are chosen to respect the Thallium EDM constraint:
\begin{eqnarray}
{\bf P0}~:~(\Phi_3,\Phi_A)&=&(0^\circ,0^\circ):
\nonumber \\ && \hspace{-2.5cm}
(M_{H_1},M_{H_2},M_{H_3};\Gamma_{H_1},\Gamma_{H_2},\Gamma_{H_3})=
(119.2,123.7,125.6;1.42,2.95,1.50)\,{\rm GeV},
\nonumber \\
{\bf P1}~:~(\Phi_3,\Phi_A)&=&(-55^\circ,30^\circ):
\nonumber \\ && \hspace{-2.5cm}
(M_{H_1},M_{H_2},M_{H_3};\Gamma_{H_1},\Gamma_{H_2},\Gamma_{H_3})=
(118.9,122.9,124.6;1.57,3.45,2.60)\,{\rm GeV},
\nonumber \\
{\bf P2}~:~(\Phi_3,\Phi_A)&=&(-80^\circ,60^\circ):
\nonumber \\ && \hspace{-2.5cm}
(M_{H_1},M_{H_2},M_{H_3};\Gamma_{H_1},\Gamma_{H_2},\Gamma_{H_3})=
(118.6,121.1,123.5;2.17,4.77,4.45)\,{\rm GeV},
\nonumber \\
{\bf P3}~:~(\Phi_3,\Phi_A)&=&(-80^\circ,90^\circ):
\nonumber \\ && \hspace{-2.5cm}
(M_{H_1},M_{H_2},M_{H_3};\Gamma_{H_1},\Gamma_{H_2},\Gamma_{H_3})=
(119.0,119.5,122.9;2.57,5.70,5.63)\,{\rm GeV},
\nonumber \\
\label{eq:p03}
\end{eqnarray}
where the masses and widths of the neutral Higgs bosons are
calculated using {\tt CPsuperH}~\cite{CPsuperH}.
We observe that the three neutral Higgs bosons are almost degenerate
with masses around 
120~GeV, and large widths comparable to the mass differences.
At the CP-conserving point {\bf P0}, the second lightest Higgs boson 
$H_2$ 
is CP odd and the CP-even $H_1$ and $H_3$ have strong {\it two-way} 
mixing. But, in the presence of non-vanishing CP phases, as at points 
{\bf 
P1}, {\bf P2} and {\bf P3},
all the three neutral Higgs bosons mix significantly. The three 
CP-violating points {\bf P1}, {\bf P2} and {\bf P3} are chosen to lie 
along the narrow region filled with black squares 
in the upper-left panel of Fig.~\ref{fig:p3pa} where the two 
contributions from $(d_e/e)^H$ and $C_S$ to the Thallium 
EDM cancel approximately and we have $|\hat{d}_{\rm Tl}| < 1\,e\,cm$. 
Thus, this selection of points complements the information available 
from 
low-energy EDM experiments.

\subsection{Backgrounds}

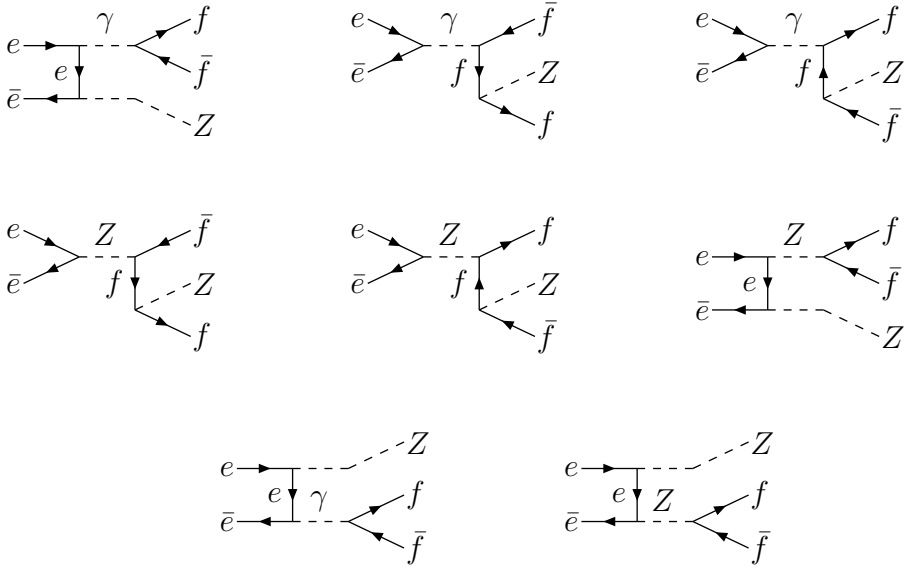
\begin{figure}[t]
\vspace*{1.5cm}
\begin{center}
{
\begin{picture}(95,79)(0,0)
\Text(15.0,60.0)[r]{$e$}
\ArrowLine(16.0,60.0)(37.0,60.0) 
\Text(47.0,65.0)[b]{$\gamma$}
\DashLine(37.0,60.0)(58.0,60.0){3.0} 
\Text(80.0,70.0)[l]{$f$}
\ArrowLine(58.0,60.0)(79.0,70.0) 
\Text(80.0,50.0)[l]{$\bar{f}$}
\ArrowLine(79.0,50.0)(58.0,60.0) 
\Text(33.0,50.0)[r]{$e$}
\ArrowLine(37.0,60.0)(37.0,40.0) 
\Text(15.0,40.0)[r]{$\bar{e}$}
\ArrowLine(37.0,40.0)(16.0,40.0) 
\DashLine(37.0,40.0)(58.0,40.0){3.0} 
\Text(80.0,30.0)[l]{$Z$}
\DashLine(58.0,40.0)(79.0,30.0){3.0} 
\end{picture} \ 
{} \qquad\allowbreak
\begin{picture}(95,79)(0,0)
\Text(15.0,70.0)[r]{$e$}
\ArrowLine(16.0,70.0)(37.0,60.0) 
\Text(15.0,50.0)[r]{$\bar{e}$}
\ArrowLine(37.0,60.0)(16.0,50.0) 
\Text(47.0,65.0)[b]{$\gamma$}
\DashLine(37.0,60.0)(58.0,60.0){3.0} 
\Text(80.0,70.0)[l]{$\bar{f}$}
\ArrowLine(79.0,70.0)(58.0,60.0) 
\Text(54.0,50.0)[r]{$f$}
\ArrowLine(58.0,60.0)(58.0,40.0) 
\Text(80.0,50.0)[l]{$Z$}
\DashLine(58.0,40.0)(79.0,50.0){3.0} 
\Text(80.0,30.0)[l]{$f$}
\ArrowLine(58.0,40.0)(79.0,30.0) 
\end{picture} \ 
{} \qquad\allowbreak
\begin{picture}(95,79)(0,0)
\Text(15.0,70.0)[r]{$e$}
\ArrowLine(16.0,70.0)(37.0,60.0) 
\Text(15.0,50.0)[r]{$\bar{e}$}
\ArrowLine(37.0,60.0)(16.0,50.0) 
\Text(47.0,65.0)[b]{$\gamma$}
\DashLine(37.0,60.0)(58.0,60.0){3.0} 
\Text(80.0,70.0)[l]{$f$}
\ArrowLine(58.0,60.0)(79.0,70.0) 
\Text(54.0,50.0)[r]{$f$}
\ArrowLine(58.0,40.0)(58.0,60.0) 
\Text(80.0,50.0)[l]{$Z$}
\DashLine(58.0,40.0)(79.0,50.0){3.0} 
\Text(80.0,30.0)[l]{$\bar{f}$}
\ArrowLine(79.0,30.0)(58.0,40.0) 
\end{picture} \ 
{} \qquad\allowbreak
\begin{picture}(95,79)(0,0)
\Text(15.0,70.0)[r]{$e$}
\ArrowLine(16.0,70.0)(37.0,60.0) 
\Text(15.0,50.0)[r]{$\bar{e}$}
\ArrowLine(37.0,60.0)(16.0,50.0) 
\Text(47.0,65.0)[b]{$Z$}
\DashLine(37.0,60.0)(58.0,60.0){3.0} 
\Text(80.0,70.0)[l]{$\bar{f}$}
\ArrowLine(79.0,70.0)(58.0,60.0) 
\Text(54.0,50.0)[r]{$f$}
\ArrowLine(58.0,60.0)(58.0,40.0) 
\Text(80.0,50.0)[l]{$Z$}
\DashLine(58.0,40.0)(79.0,50.0){3.0} 
\Text(80.0,30.0)[l]{$f$}
\ArrowLine(58.0,40.0)(79.0,30.0) 
\end{picture} \ 
{} \qquad\allowbreak
\begin{picture}(95,79)(0,0)
\Text(15.0,70.0)[r]{$e$}
\ArrowLine(16.0,70.0)(37.0,60.0) 
\Text(15.0,50.0)[r]{$\bar{e}$}
\ArrowLine(37.0,60.0)(16.0,50.0) 
\Text(47.0,65.0)[b]{$Z$}
\DashLine(37.0,60.0)(58.0,60.0){3.0} 
\Text(80.0,70.0)[l]{$f$}
\ArrowLine(58.0,60.0)(79.0,70.0) 
\Text(54.0,50.0)[r]{$f$}
\ArrowLine(58.0,40.0)(58.0,60.0) 
\Text(80.0,50.0)[l]{$Z$}
\DashLine(58.0,40.0)(79.0,50.0){3.0} 
\Text(80.0,30.0)[l]{$\bar{f}$}
\ArrowLine(79.0,30.0)(58.0,40.0) 
\end{picture} \ 
{} \qquad\allowbreak
\begin{picture}(95,79)(0,0)
\Text(15.0,60.0)[r]{$e$}
\ArrowLine(16.0,60.0)(37.0,60.0) 
\Text(47.0,65.0)[b]{$Z$}
\DashLine(37.0,60.0)(58.0,60.0){3.0} 
\Text(80.0,70.0)[l]{$f$}
\ArrowLine(58.0,60.0)(79.0,70.0) 
\Text(80.0,50.0)[l]{$\bar{f}$}
\ArrowLine(79.0,50.0)(58.0,60.0) 
\Text(33.0,50.0)[r]{$e$}
\ArrowLine(37.0,60.0)(37.0,40.0) 
\Text(15.0,40.0)[r]{$\bar{e}$}
\ArrowLine(37.0,40.0)(16.0,40.0) 
\DashLine(37.0,40.0)(58.0,40.0){3.0} 
\Text(80.0,30.0)[l]{$Z$}
\DashLine(58.0,40.0)(79.0,30.0){3.0} 
\end{picture} \ 
{} \qquad\allowbreak
\begin{picture}(95,79)(0,0)
\Text(15.0,60.0)[r]{$e$}
\ArrowLine(16.0,60.0)(37.0,60.0) 
\DashLine(37.0,60.0)(58.0,60.0){3.0} 
\Text(80.0,70.0)[l]{$Z$}
\DashLine(58.0,60.0)(79.0,70.0){3.0} 
\Text(33.0,50.0)[r]{$e$}
\ArrowLine(37.0,60.0)(37.0,40.0) 
\Text(15.0,40.0)[r]{$\bar{e}$}
\ArrowLine(37.0,40.0)(16.0,40.0) 
\Text(47.0,44.0)[b]{$\gamma$}
\DashLine(37.0,40.0)(58.0,40.0){3.0} 
\Text(80.0,50.0)[l]{$f$}
\ArrowLine(58.0,40.0)(79.0,50.0) 
\Text(80.0,30.0)[l]{$\bar{f}$}
\ArrowLine(79.0,30.0)(58.0,40.0) 
\end{picture} \ 
{} \qquad\allowbreak
\begin{picture}(95,79)(0,0)
\Text(15.0,60.0)[r]{$e$}
\ArrowLine(16.0,60.0)(37.0,60.0) 
\DashLine(37.0,60.0)(58.0,60.0){3.0} 
\Text(80.0,70.0)[l]{$Z$}
\DashLine(58.0,60.0)(79.0,70.0){3.0} 
\Text(33.0,50.0)[r]{$e$}
\ArrowLine(37.0,60.0)(37.0,40.0) 
\Text(15.0,40.0)[r]{$\bar{e}$}
\ArrowLine(37.0,40.0)(16.0,40.0) 
\Text(47.0,44.0)[b]{$Z$}
\DashLine(37.0,40.0)(58.0,40.0){3.0} 
\Text(80.0,50.0)[l]{$f$}
\ArrowLine(58.0,40.0)(79.0,50.0) 
\Text(80.0,30.0)[l]{$\bar{f}$}
\ArrowLine(79.0,30.0)(58.0,40.0) 
\end{picture} \ 
}
\end{center}
\vspace{-1.0cm}
\noindent
\caption{\it Diagrams for the SM background
process $e^+e^-\rightarrow f\bar{f}\,Z$ with $f=b$ and $\tau^-$.}
\label{fig:bkg_diag}
\end{figure}

\begin{figure}[t]
\vspace{-1.5cm}
\centerline{\epsfig{figure=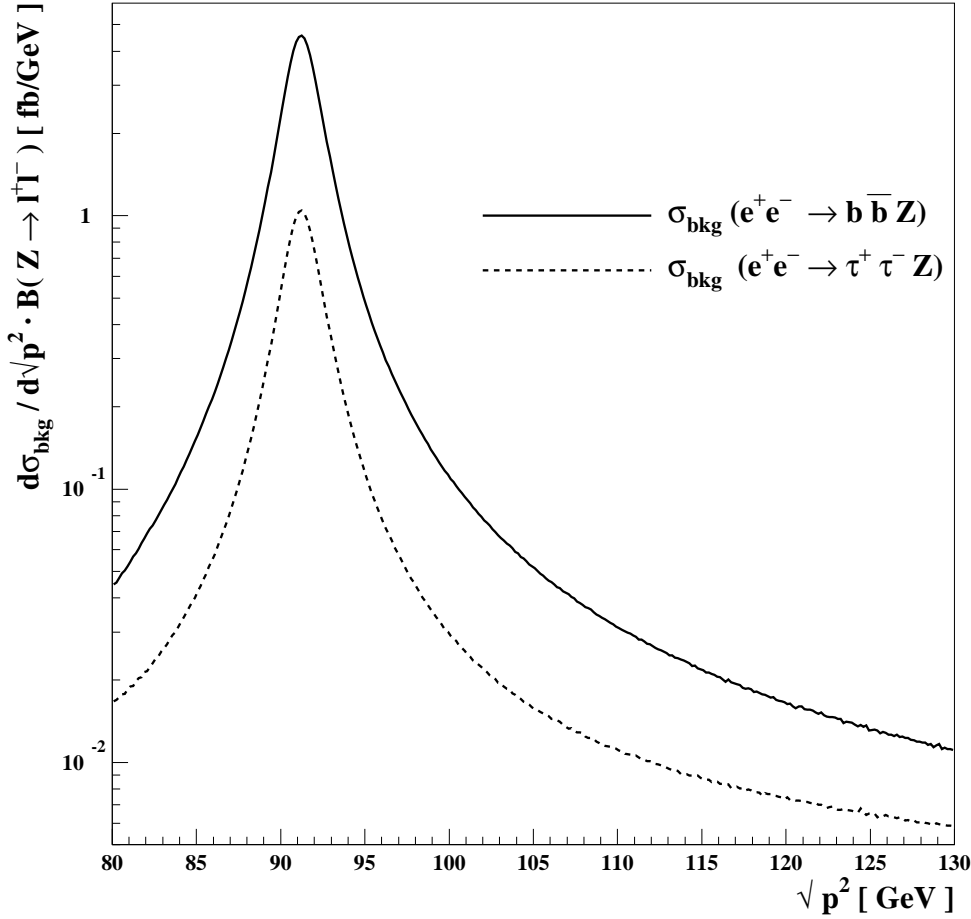,height=14cm,width=14cm}}
\caption{{\it 
The product of the differential
background cross section and the branching fraction of 
the $Z$ boson into electrons and muons, 
$\frac{d\sigma_{\rm bkg}}{d\sqrt{p^2}}\times B(Z\rightarrow l^+l^-)$, 
as
a a function of $\sqrt{p^2}$  in units of fb/GeV. 
The solid line is for the process
$e^+e^-\rightarrow b \bar{b} Z$ and the dashed line for 
$e^+e^-\rightarrow \tau^- \tau^+ Z$.
}}
\label{fig:bkg}
\end{figure}

We consider the two final states $f=b$ and $f=\tau^-$ for the 
Higgs-boson
decays. Therefore, before showing the numerical results for the 
Higgsstrahlung processes, we first consider the SM background processes
$e^+e^-\rightarrow f \bar{f} Z$ with $f=b$ and $\tau^-$, omitting the
Higgs-mediated diagrams, as
seen in Fig.~\ref{fig:bkg_diag}. The background cross sections are 
evaluated using {\tt CompHEP} \cite{CompHEP}. 
We show in Fig.~\ref{fig:bkg} the product of the differential 
background
cross section
${d\sigma_{\rm bkg}}/{d\sqrt{p^2}}$ and the branching fraction of 
the $Z$ boson into electrons and muons, $B(Z\rightarrow l^+l^-)$, as 
a function
of the invariant mass of the fermion-antifermion pair $\sqrt{p^2}$ in 
units of fb/GeV. The solid line is for $f=b$ and the dashed line for
$f=\tau^-$. We used for the leptonic branching fraction~\cite{PDG}:
\begin{equation}
B(Z\rightarrow l^+l^-)\ =\
B(Z\rightarrow e^+e^-)+
B(Z\rightarrow \mu^+\mu^-)\ =\ 6.73\times 10^{-2}\; .
\end{equation}
We  note that  the product of  the  background  cross section and  the
branching fraction of the $Z$ boson into light leptons is smaller than
$\sim 0.03\,(0.01)$ fb/GeV for $f=b\,(\tau^-)$ when $\sqrt{p^2} > 110$
GeV.

\subsection{CP-Conserving Cross Sections}

\begin{figure}[t]
\vspace{-1.5cm}
\centerline{\epsfig{figure=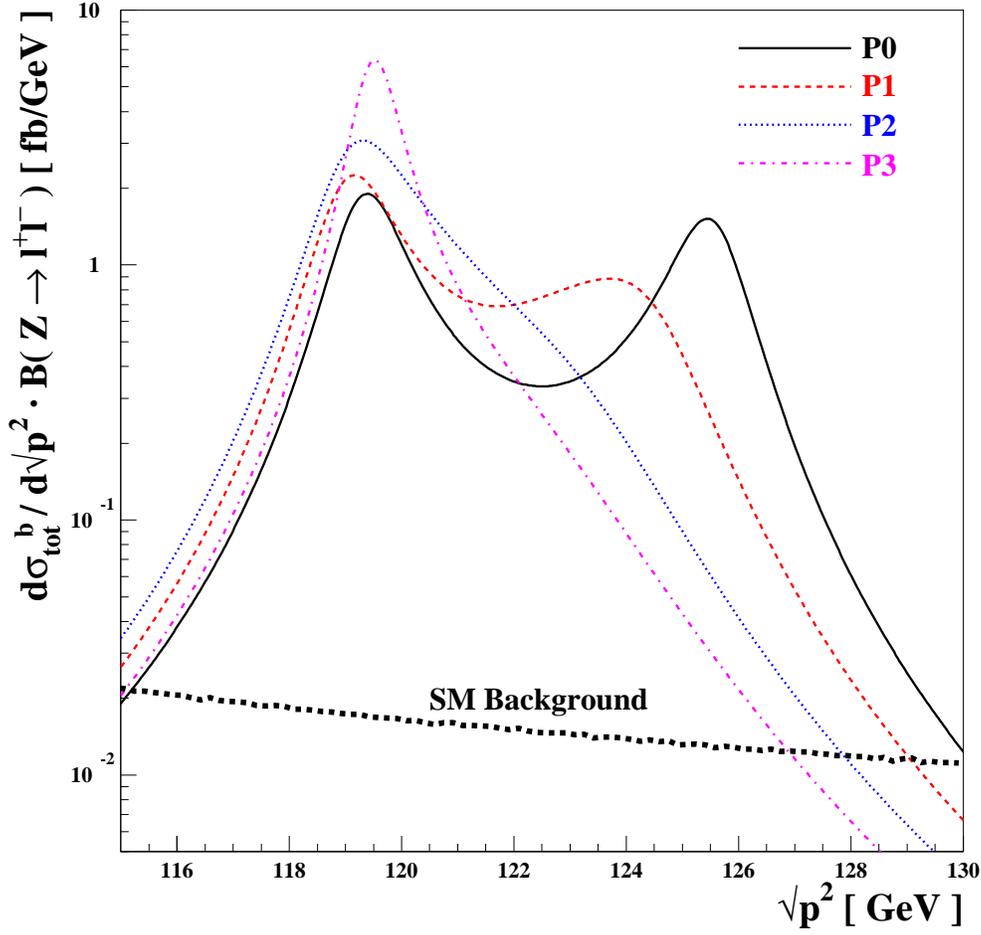,height=14cm,width=14cm}}
\vspace{-0.5 cm}
\caption{{\it  The   differential  total  cross section  multiplied by
$B(Z\rightarrow l^+l^-)$ when the produced Higgs bosons decay into $b$
quarks:  $\frac{d\sigma^b_{\rm  tot}}{d\sqrt{p^2}}\times B(Z\rightarrow
l^+l^-)$.  The CP-conserving     two-way mixing ({\bf  P0})    and the
CP-violating tri-mixing ({\bf P1}-{\bf P3}) scenarios have been taken,
see~(\ref{eq:tri}) and (\ref{eq:p03}). The solid line is for {\bf P0},
the dashed  line for {\bf P1},  the dotted line  for {\bf P2}, and the
dash-dotted line for {\bf P3}.  The  SM background cross section  from
Fig.~\ref{fig:bkg} is also shown.  }}
\label{fig:bquark}
\end{figure}

The  differential total cross section can  be obtained by summing over
the polarizations $P$ and $\bar{P}$ in~(\ref{eq:cx}). We have
\begin{equation}
\frac{d\sigma_{\rm tot}^f}{d\sqrt{p^2}}\ =\
\frac{8\,\hat\sigma_1^f(p^2)}{\sqrt{p^2}}\,,
\end{equation}
where $\hat\sigma_1^f$ is defined as in~(\ref{eq:s14}).
We show in  Fig.~\ref{fig:bquark} the differential total cross section
multiplied by $B(Z\rightarrow l^+l^-)$ when  the produced Higgs bosons
decay into  $b$ quarks: $\frac{d\sigma^b_{\rm tot}}{d\sqrt{p^2}}\times
B(Z\rightarrow l^+l^-)$.  The  cross sections are significantly larger
(0.1-5 fb/GeV) than that of the SM background ($< 0.02$ fb/GeV) around
the peaks.  In  the CP-conserving case {\bf   P0}, we clearly  see two
peaks of the CP-even Higgs bosons at $\sqrt{p^2}=119.2$ GeV ($H_1$)  
and at $\sqrt{p^2}=125.6$ GeV ($H_3$), see  also (\ref{eq:p03}). 
The second lightest CP-odd $H_2$ does not contribute.
However,  in the CP-violating cases
this  two-peak structure  becomes  less  clear  as the phase  $\Phi_A$
increases: see the dashed ({\bf P1} : $\Phi_A=30^\circ$), dotted ({\bf
P2}     :   $\Phi_A=60^\circ$),  and      dashed-dotted   ({\bf P3}  :
$\Phi_A=90^\circ$) lines.
The disappearance of the two-peak structure is the combined effect of 
increasing (decreasing) $H_2$ ($H_3$) coupling to $Z$ bosons,
$g_{H_2VV}^2$ ($g_{H_3VV}^2$), and decreasing mass difference between 
$H_1$ and $H_2$ without visible changes in $M_{H_1}$ around 119 GeV.
The sensitivity of this  CP-conserving  quantity to the  CP-violating
phases will be measurable by examining the production lineshape at the
ILC. For  example, an integrated  luminosity  larger than  $\sim$  100
fb$^{-1}$  would enable a  difference  of $\sim$  0.1 fb/GeV  in cross
sections to be distinguished easily.

As was emphasized in \cite{ELP1}, the resonance lineshape of a coupled
system of neutral Higgs  bosons is not a process-independent quantity,
but  crucially  depends  on  its  production and  decay  channels.   A
combined analysis  of the different  production and decay  channels at
the LHC, ILC and  PLC can shed light on whether one  is dealing with a
single,  two- or  multi-component  system of  Higgs  bosons.  Such  an
extensive analysis is beyond the scope of the present paper and may be
given elsewhere. As we demonstrate explicitly below,
the possible observation of non-zero CP asymmetries could give further
insight into the CP composition of such a resonant Higgs boson system
\footnote{We  should  note  that  performing  an overall  fit  to  the
production lineshape becomes more challenging in the presence of
CP  violation.  Specifically, one  has three  masses of  neutral Higgs
bosons, six widths including  off-diagonal absorptive effects, and two
independent Higgs-boson  couplings to $Z$  bosons.  
Therefore, the analysis of other observables in addition to the total
cross section would  be very useful for the  complete determination of
the parameters.}.

\begin{figure}[t]
\vspace{-1.0cm}
\centerline{\epsfig{figure=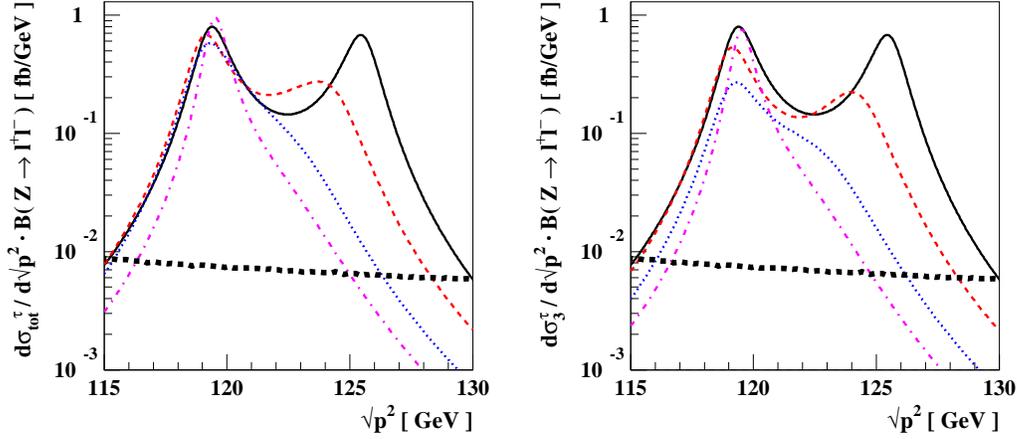,height=14cm,width=14cm}}
\vspace{-6.9 cm}
\caption{{\it Two differential CP-conserving 
cross sections (multiplied by $B(Z\rightarrow l^+l^-)$) that are 
observable when the produced Higgs bosons decay into $\tau$ leptons: 
$\frac{d\sigma^\tau_{\rm tot}}{d\sqrt{p^2}}\times B(Z\rightarrow 
l^+l^-)$ 
(left panel) and $\frac{d\sigma^\tau_3}{d\sqrt{p^2}}\times 
B(Z\rightarrow 
l^+l^-)$ (right panel). The line styles are as in 
Fig.~\ref{fig:bquark}.
}}
\label{fig:tau}
\end{figure}

When the  produced Higgs  bosons  decay  into  $\tau$ leptons, we  can
construct another CP-conserving cross section in addition to the total
cross section,  by measuring  the    transverse polarizations of   tau
leptons  or, equivalently, by examining  the polar and azimuthal angle
distributions of the  charged  hadrons coming from   the $\tau$-lepton
decays:
\begin{equation}
\frac{d\sigma_3^\tau}{d\sqrt{p^2}}\ =\
\frac{8\,\hat\sigma_3^\tau(p^2)}{\sqrt{p^2}}\,.
\end{equation}
This   is  related  to   the  polarization  coefficient   $C_3^f(p^2)$
in~(\ref{eq:cx})  and (\ref{eq:taudecay}). The CP-conserving total and
transverse  differential cross sections multiplied by $B(Z\rightarrow
l^+l^-)$ are shown in the left and right panels of Fig.~\ref{fig:tau},
respectively.   The cross sections    are  smaller than  that   of the
$b$-quark  case ($\sim$  0.01--1  fb/GeV),  but the transverse   cross
section $d\sigma^\tau_3/d\sqrt{p^2}$ provides extra sensitivity to the
CP-violating  phases in the $\tau$-lepton   case,  in addition to  the
total cross section.

\subsection{CP-Violating Cross Sections and Asymmetries}

When the  produced Higgs bosons  decay into $\tau$  leptons, there are
two  CP-violating  cross  sections   which    are defined  using   the
longitudinal and transverse polarizations of $\tau$ leptons:
\begin{equation}
\frac{d\Delta\sigma_L^\tau}{d\sqrt{p^2}}\ =\
\frac{8\,\hat\sigma_2^\tau(p^2)}{\sqrt{p^2}}\ , \qquad\qquad
\frac{d\Delta\sigma_T^\tau}{d\sqrt{p^2}}\ =\
\frac{8\,\hat\sigma_4^\tau(p^2)}{\sqrt{p^2}}\ ,
\end{equation}
where $\Delta\sigma_L^\tau$ and  $\Delta\sigma_T^\tau$ are  related to
the   polarization  coefficients      $C_2^f(p^2)$ and   $C_4^f(p^2)$,
respectively.  The    polarization  coefficients   $C_2^f(p^2)$    and
$C_4^f(p^2)$  can  be  determined by  measuring  the  longitudinal and
transverse   polarizations   of   $\tau$       leptons,   respectively
[cf.~(\ref{eq:cx})    and   (\ref{eq:taudecay})].      We   define the
corresponding longitudinal and transverse CP asymmetries as follows:
\begin{equation}
a_L^\tau(p^2)\ \equiv\ \frac{{d\Delta\sigma_L^\tau}/{d\sqrt{p^2}}}
                    {{d\sigma_{\rm tot}^\tau}/{d\sqrt{p^2}}}\ =\
\frac{\sigma_2^\tau(p^2)}{\sigma_1^\tau(p^2)}\ , \qquad
a_T^\tau(p^2)\ \equiv\ \frac{{d\Delta\sigma_T^\tau}/{d\sqrt{p^2}}}
                    {{d\sigma_{\rm tot}^\tau}/{d\sqrt{p^2}}}\ =\
\frac{\sigma_4^\tau(p^2)}{\sigma_1^\tau(p^2)}\ .
\end{equation}

\begin{figure}[t]
\vspace{-1.0cm}
\centerline{\epsfig{figure=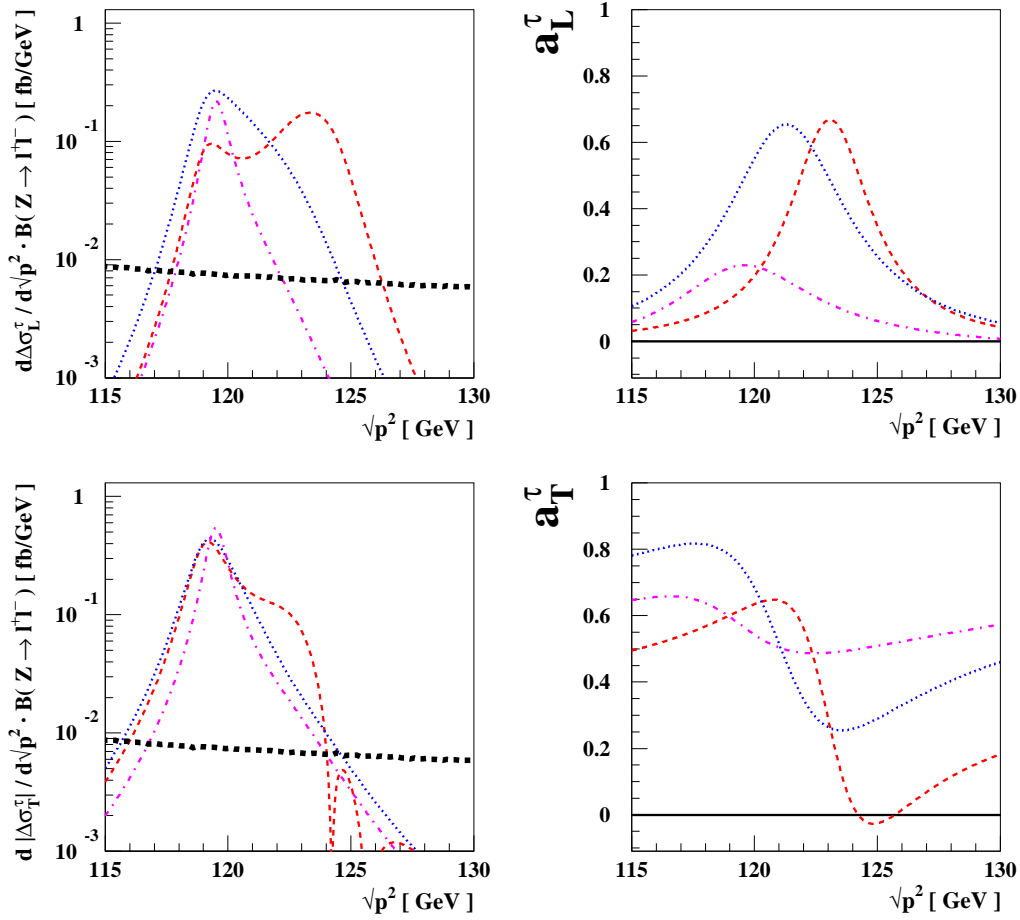,height=14cm,width=14cm}}
\vspace{-1.0 cm}
\caption{{\it Two differential CP-violating 
cross sections (multiplied by $B(Z\rightarrow l^+l^-)$) observable when 
the produced Higgs bosons decay into $\tau$ leptons: 
$\frac{d\Delta\sigma^\tau_L}{d\sqrt{p^2}}\times B(Z\rightarrow l^+l^-)$ 
(upper-left panel) and 
$\frac{d\Delta\sigma^\tau_T}{d\sqrt{p^2}}\times B(Z\rightarrow l^+l^-)$
(lower-left panel). The two 
corresponding CP-violating asymmetries are shown in the right column:
$a^\tau_L$ (upper-right panel) and $a^\tau_T$ (lower-right panel).
The lines styles are the same as in Fig.~\ref{fig:bquark}.
}}
\label{fig:tau_cp}
\end{figure}

We show in Fig.~\ref{fig:tau_cp} the CP-violating cross sections (left
column)  and CP  asymmetries (right  column).  The  CP-violating cross
sections  are large  enough  to be  measured,  assuming luminosity  of
$>100\,{\rm fb}^{-1}$, for each choice of the CP-violating phases, and
the CP asymmetry can be as  large as 80\%.  Analysis of the production
lineshape would enable the cases with different CP-violating phases to
be  distinguished from  each other.  Since the  CP-violating scenarios
chosen  respect the  low-energy EDM  constraints, these  examples show
that  linear-collider   measurements  are  complementary,   and  large
CP-violating effects cannot be excluded {\it a priori}.

\section{Conclusions}
\label{sec:conclusion}

We have shown that the Higgsstrahlung process $e^+e^- \rightarrow
H_{1,2,3} (Z \to e^+e^-,\mu^+\mu^-)$, with subsequent decays $H_{1,2,3}
\to b \bar{b},\ \tau^-\tau^+$ is potentially a useful channel of 
searching
for radiative Higgs-sector CP violation in the MSSM. The recoil-mass
method would enable one to measure the invariant mass of a $b \bar{b}$ 
or
$\tau^-\tau^+$ pair produced in Higgs decay with a precision as good as
1 GeV.  In tri-mixing scenarios where all three neutral Higgs
states of the MSSM are nearly degenerate and mix significantly, this
accuracy would enable details of complicated lineshapes to be
disentangled.  An integrated luminosity of 100 fb$^{-1}$ would already 
be
sufficient to measure CP-violating parameters via their influences on 
the
production lineshape of a coupled system of neutral Higgs bosons 
decaying
into $b\bar{b}$ quarks. Measurements of the Higgs bosons decaying into
$\tau^-\tau^+$ would enable an additional CP-conserving cross section 
and
two CP-violating asymmetries to be defined in terms of the longitudinal
and transverse polarizations of the $\tau$ leptons.

We  find that these  CP asymmetries  could be  as large  as 80\%, even
after  taking into account  the constraints from the Thallium electric
dipole moment, and different CP-violating  models compatible with  the
Thallium  data  could  be  distinguished.   Thus,  measurements  of CP
violation  in  Higgsstrahlung  at an $e^+e^-$   linear  collider would
complement high-precision low-energy measurements, and might provide a
signal, even if none is visible in the low-energy experiments.

Finally, we should stress that the analysis presented in this paper is
general and   applies  equally  well to  extended   CP-violating Higgs
sectors with similar phenomenological  features.  CP~violation and its
relation  to the   cosmological     baryon asymmetry are   among   the
outstanding questions in possible physics  beyond the Standard  Model,
and measurements of the Higgsstrahlung  process could provide a unique
window that could shine valuable light on attempts to relate these two
puzzles.

\subsection*{Acknowledgements}
The  work of JSL and  AP is supported  in  part by  the PPARC research
grant PPA/G/O/2002/00471.

\newpage



\begin{thebibliography}{99}

\bibitem{HS} J.~R.~Ellis, M.~K.~Gaillard and D.~V.~Nanopoulos,
  Nucl.\ Phys.\ B {\bf 106} (1976) 292;
B.~L.~Ioffe and V.~A.~Khoze, Sov.\ J.\ Part.\ Nucl.\ {\bf 9} (1978) 50
[Fiz.\ Elem.\ Chast.\ Atom.\ Yadra {\bf 9} (1978) 118];
B.~W.~Lee, C.~Quigg and H.~B.~Thacker, Phys.\ Rev.\ D {\bf 16} (1977) 
1519.

\bibitem{LEPHWG}
  R.~Barate {\it et al.}  [ALEPH, DELPHI, L3 and OPAL Collaborations],
  {\it Search for the standard model Higgs boson at LEP},
  Phys.\ Lett.\ B {\bf 565} (2003) 61
  [arXiv:hep-ex/0306033];
see also {\tt http://lephiggs.web.cern.ch/LEPHIGGS/papers/index.html}.

\bibitem{LEPEWWG}
The ALEPH, DELPHI, L3 and OPAL Collaborations, the LEP Electroweak 
Working 
Group, the 
SLD Electroweak and Heavy Flavour Groups,
{\it A 
combination of preliminary electroweak measurements and constraints  
on
the standard model},
  arXiv:hep-ex/0312023;
as updated on {\tt http://lepewwg.web.cern.ch/LEPEWWG/Welcome.html}.

\bibitem{HPN} For reviews, see, H.P. Nilles, Phys.\ Rep.\ {\bf 110}
  (1984) 1; H. Haber and G.  Kane, Phys.\ Rep.\ {\bf 117} (1985) 75;
  J.F.  Gunion, H.E. Haber, G.L.  Kane and S. Dawson, {\it The Higgs
  Hunter's Guide}, (Addison-Wesley, Reading, MA, 1990).

\bibitem{APLB} A. Pilaftsis, Phys.\ Rev.\ D {\bf 58} (1998) 096010; 
  Phys.\ Lett.\ B {\bf 435} (1998) 88.

\bibitem{PW} A. Pilaftsis and  C.E.M. Wagner, Nucl.\ Phys.\ B {\bf 553}
  (1999) 3.

\bibitem{Demir} D.A. Demir, Phys.\ Rev.\ D {\bf 60} (1999) 055006.

\bibitem{CDL} S.Y. Choi, M. Drees and J.S. Lee, Phys.\ Lett.\ B {\bf
  481} (2000) 57.

\bibitem{CEPW} M. Carena, J. Ellis, A. Pilaftsis and C.E.M. Wagner,
  Nucl.\ Phys.\ B {\bf 586} (2000) 92.

\bibitem{CPX} 
M.~Carena, J.~R.~Ellis, A.~Pilaftsis and C.~E.~M.~Wagner,
  Phys.\ Lett.\ B {\bf 495} (2000) 155
  [arXiv:hep-ph/0009212].

\bibitem{CEMPW}
M.~Carena, J.~R.~Ellis, S.~Mrenna, A.~Pilaftsis and C.~E.~M.~Wagner,
  Nucl.\ Phys.\ B {\bf 659} (2003) 145
  [arXiv:hep-ph/0211467].


\bibitem{APNPB} A.~Pilaftsis, Nucl.\ Phys.\ B {\bf 504} (1997) 61.

\bibitem{ELP1} 
J.~R.~Ellis, J.~S.~Lee and A.~Pilaftsis,
Phys.\ Rev.\ D {\bf 70} (2004) 075010;
%
J.S. Lee, hep-ph/0409020.

\bibitem{ELP2}
  J.~R.~Ellis, J.~S.~Lee and A.~Pilaftsis,
  Nucl.\ Phys.\ B {\bf 718} (2005) 247.

\bibitem{ELP3}
  J.~R.~Ellis, J.~S.~Lee and A.~Pilaftsis,
  Phys.\ Rev.\ D {\bf 71} (2005) 075007.


\bibitem{CPdirect} S.~Y.~Choi, J.~Kalinowski, G.~Moortgat-Pick and
  P.~M.~Zerwas, Eur.\ Phys.\ J.\ C {\bf 22} (2001) 563; S.~Y.~Choi,
  A.~Djouadi, M.~Guchait, J.~Kalinowski, H.~S.~Song and P.~M.~Zerwas,
  Eur.\ Phys.\ J.\ C {\bf 14} (2000) 535; 
  A.~Bartl, H.~Fraas, O.~Kittel and W.~Majerotto, Phys.\
  Rev.\ D {\bf 69} (2004) 035007;
%
S.~Y.~Choi, M.~Drees and B.~Gaissmaier,
Phys.\ Rev.\ D {\bf 70} (2004) 014010.

\bibitem{CPsoft1}
A.~Bartl, S.~Hesselbach, K.~Hidaka, T.~Kernreiter and W.~Porod, 
Phys.\ Lett.\ B {\bf 573} (2003) 153; 
Phys.\ Rev.\ D {\bf 70} (2004) 035003;
%
S.~Y.~Choi,
Phys.\ Rev.\ D {\bf 69} (2004) 096003;
%
S.~Y.~Choi, M.~Drees, B.~Gaissmaier and J.~Song,
Phys.\ Rev.\ D {\bf 69} (2004) 035008;
%
S.~Y.~Choi and Y.~G.~Kim,
Phys.\ Rev.\ D {\bf 69} (2004) 015011;
%
J.~A.~Aguilar-Saavedra,
Phys.\ Lett.\ B {\bf 596} (2004) 247;
%
Nucl.\ Phys.\ B {\bf 697} (2004) 207;
%
T.~Gajdosik, R.~M.~Godbole and S.~Kraml,
JHEP {\bf 0409} (2004) 051.
%

\bibitem{CPsoft2} For an extensive review, see D.~J.~H.~Chung,
L.~L.~Everett, G.~L.~Kane, S.~F.~King, J.~Lykken and L.~T.~Wang,
hep-ph/0312378. 


\bibitem{EDM1}  J. Ellis,   S.  Ferrara  and D.V. Nanopoulos,   Phys.\
  Lett.\  B {\bf 114} (1982)  231; W.  Buchm\"uller  and D. Wyler, 
Phys.\
  Lett.\ B {\bf   121} (1983) 321;  J. Polchinski   and M.   Wise, 
Phys.\
  Lett.\ B {\bf 125} (1983)  393; F. del Aguila,  M.  Gavela, J.  
Grifols
  and  A.   Mendez, Phys.\ Lett.\ B {\bf 126}  (1983) 71; M.   Dugan, 
B.
  Grinstein and L.   Hall, Nucl.\   Phys.\  B {\bf 255}  (1985)  413;  
R.
  Garisto and J.D.  Wells, Phys.\ Rev.\ D {\bf 55} (1997) 1611.

\bibitem{EDM2} T.  Ibrahim and P.  Nath, Phys.\ Rev.\ D {\bf 58}
  (1998) 111301; Phys.\ Rev.\ D {\bf 61} (2000)~093004; M.~Brhlik,
  L.~Everett, G.L.~Kane and J.~Lykken, Phys.\ Rev.\ Lett.\ {\bf 83}
  (1999) 2124; Phys.\ Rev.\ D {\bf 62} (2000) 035005; S.~Pokorski,
  J.~Rosiek and C.A.~Savoy, Nucl.\ Phys.\ B {\bf 570} (2000) 81;
  E.~Accomando, R.~Arnowitt and B.~Dutta, Phys.\ Rev.\ D {\bf 61} 
(2000)
  115003; A.~Bartl, T.~Gajdosik, W.~Porod, P.~Stockinger and
  H.~Stremnitzer, Phys.\ Rev.\ D {\bf 60} (1999) 073003; T.~Falk,
  K.A.~Olive, M.~Pospelov and R.~Roiban, Nucl.\ Phys.\ B {\bf 560} 
(1999) 3. 

\bibitem{CKP} For two-loop Higgs-mediated contributions to EDMs in the
  CP-violating MSSM, see D.  Chang, W.-Y. Keung and A. Pilaftsis, 
Phys.\
  Rev.\ Lett.\ {\bf 82} (1999) 900; A. Pilaftsis, Nucl.\ Phys.\ B {\bf
  644} (2002) 263; D.~A.~Demir, O.~Lebedev, K.~A.~Olive, M.~Pospelov 
and
  A.~Ritz, Nucl.\ Phys.\ B {\bf 680} (2004) 339;
  K.~A.~Olive, M.~Pospelov, A.~Ritz and Y.~Santoso,
  arXiv:hep-ph/0506106.


\bibitem{Bmeson1} 
  P.H.  Chankowski and L. Slawianowska, Phys.\ Rev.\ D {\bf 63} (2001)
  054012;   C.S.  Huang, W.  Liao,  Q.-S.   Yan  and S.-H.  Zhu, Phys.\
  Rev.\ D {\bf 63} (2001)  114021;  D.~A.~Demir and K.~A.~Olive,   
Phys.\
  Rev.\ D {\bf  65}  (2002) 034007; M.~Boz  and  N.~K.~Pak, Phys.\ 
Lett.\
  B {\bf 531} (2002) 119; A.~J.~Buras, P.~H.~Chankowski, J.~Rosiek and
  L.~Slawianowska, Nucl.\ Phys.\ B {\bf 659} (2003) 3;
  T.~Ibrahim and  P.~Nath, Phys.\ Rev.\ D {\bf 67}
  (2003) 016005; Phys.\  Rev.\ D {\bf 67}  (2003)  095003;
  A.~Dedes,
  Mod.\ Phys.\ Lett.\ A {\bf 18} (2003) 2627;
  M.~E.~Gomez, T.~Ibrahim, P.~Nath and S.~Skadhauge,
  hep-ph/0506243.

\bibitem{DP} A. Dedes and A. Pilaftsis, Phys.\ Rev.\ D {\bf 67} (2003)
  015012.

\bibitem{baryog}
For recent studies, see,
T.~Konstandin, T.~Prokopec and M.~G.~Schmidt,
  arXiv:hep-ph/0410135;
  K.~Funakubo, S.~Tao and F.~Toyoda,
  Prog.\ Theor.\ Phys.\  {\bf 109} (2003) 415;
  M.~Carena, M.~Quiros, M.~Seco and C.~E.~M.~Wagner,
  Nucl.\ Phys.\ B {\bf 650} (2003) 24;
J.~M.~Cline,
hep-ph/0201286.


\bibitem{INhiggs} T. Ibrahim and P. Nath, Phys.\ Rev.\ D {\bf 63}
  (2001) 035009; Phys.\ Rev.\ D {\bf 66} (2002) 015005; T. Ibrahim,
  Phys.\ Rev.\ D {\bf 64} (2001) 035009; S.~W.~Ham, S.~K.~Oh,
  E.~J.~Yoo, C.~M.~Kim and D.~Son, Phys.\ Rev.\ D {\bf 68} (2003)
  055003; T.~Ibrahim, P.~Nath and A.~Psinas,
  Phys.\ Rev.\ D {\bf 70} (2004) 035006.

\bibitem{CEPW2} M. Carena, J. Ellis, A. Pilaftsis and C.E.M. Wagner,
  Nucl.\ Phys.\ B {\bf 625} (2002) 345.

\bibitem{KW} G.L. Kane and L.-T. Wang, Phys.\ Lett.\ B {\bf 488} (2000) 
383.

\bibitem{HeinCP} S.~Heinemeyer, Eur.\ Phys.\ J. C {\bf 22} (2001) 521.

\bibitem{Maria} 
E.~Akhmetzyanova, M.~Dolgopolov and M.~Dubinin,
hep-ph/0405264;
%
I.~F.~Ginzburg and M.~Krawczyk,
hep-ph/0408011.

\bibitem{CHL} S.Y.  Choi and J.S. Lee,  Phys.\ Rev.\ D {\bf 61} (2000)
  015003; S.Y. Choi, K. Hagiwara  and J.S. Lee,  Phys.\ Rev.\ D {\bf 
64}
  (2001) 032004;   S.~Y.~Choi,  M.~Drees, J.~S.~Lee   and J.~Song, 
Eur.\
  Phys.\ J.\ C {\bf 25} (2002) 307.

\bibitem{CPpp} 
W.~Bernreuther and A.~Brandenburg,
Phys.\ Lett.\ B {\bf 314} (1993) 104;
%
W.~Bernreuther and A.~Brandenburg,
Phys.\ Rev.\ D {\bf 49} (1994) 4481;
%
  A. Dedes and S. Moretti, Phys.\ Rev.\ Lett.\ {\bf 84}
  (2000) 22; Nucl.\ Phys.\ B {\bf 576} (2000) 29; S.Y. Choi and J.S. 
Lee,
  Phys.\ Rev.\ D {\bf 61} (2000) 115002; S.Y.  Choi, K.  Hagiwara and
  J.S. Lee, Phys.\ Lett.\ B {\bf 529} (2002) 212;
  A.~Arhrib,  D.~K.~Ghosh and O.C.~Kong,
  Phys.\ Lett.\ B {\bf 537}   (2002)  217;
  E.~Christova, H.~Eberl, W.~Majerotto and S.~Kraml,
  Nucl.\ Phys.\ B {\bf 639} (2002) 263; JHEP {\bf 0212} (2002) 021;
  W.~Khater and P.~Osland, Nucl.\ Phys.\ B {\bf 661} (2003) 209;
F.~Borzumati, J.~S.~Lee and W.~Y.~Song,
production in
Phys.\ Lett.\ B {\bf 595} (2004) 347.

\bibitem{CFLMP} B.E. Cox, J.R. Forshaw, J.S. Lee, J.W. Monk and
  A.~Pilaftsis, Phys.\ Rev.\ D {\bf 68} (2003) 075004.

\bibitem{Akeroyd}
  A.G.~Akeroyd, Phys.\ Rev.\ D {\bf 68} (2003) 077701;
  D.~K.~Ghosh, R.~M.~Godbole and D.~P.~Roy,
  arXiv:hep-ph/0412193;
  D.~K.~Ghosh and S.~Moretti,
  arXiv:hep-ph/0412365.

\bibitem{KMR} V.A. Khoze, A.D. Martin and M.G. Ryskin, 
  Eur.\ Phys.\ J.\ C {\bf 34} (2004) 327.


\bibitem{CPee} B.~Grzadkowski,  J.~F.~Gunion and J.~Kalinowski, Phys.\
  Rev.\ D {\bf 60}  (1999) 075011;  A.G.~Akeroyd  and A.  Arhrib, 
Phys.\
  Rev.\ D {\bf 64} (2001) 095018.

\bibitem{CPmumu} D.~Atwood and A.~Soni, Phys.\ Rev.\ D {\bf 52} (1995)
  6271; B.~Grzadkowski and J.F.~Gunion, Phys.\ Lett.\ B {\bf 350} 
(1995)
  218; A.~Pilaftsis, Phys.\ Rev.\ Lett.\ {\bf 77} (1996) 4996;
  S.Y.~Choi and J.S.~Lee, Phys.\ Rev.\ D {\bf 61} (2000) 111702;
  E.~Asakawa, S.Y.~Choi and J.S.~Lee, Phys.\ Rev.\ D {\bf 63} (2001) 
015012;
  S.Y.~Choi, M.~Drees, B.~Gaissmaier and J.S.~Lee, Phys.\ Rev.\ D {\bf 
64}
  (2001) 095009; M.S.~Berger, Phys.\ Rev.\ Lett.\ {\bf 87} (2001) 
131801;
  C.~Blochinger {\it et al.}, hep-ph/0202199.

\bibitem{CPphoton} 
  S.~Y.~Choi and J.~S.~Lee, Phys.\ Rev.\ D {\bf 62}
  (2000) 036005; 
  J.~S.~Lee, hep-ph/0106327;
  S.~Y.~Choi, B.~C.~Chung, P.~Ko and J.~S.~Lee, Phys.\ Rev.\ D {\bf 66}
  (2002) 016009; 
  S.~Y.~Choi, J.~Kalinowski, J.~S.~Lee, M.~M.~Muhlleitner, M.~Spira and
P.~M.~Zerwas,
  Phys.\ Lett.\ B {\bf 606} (2005) 164;
  S.~Heinemeyer and M.~Velasco,
  hep-ph/0506267.

\bibitem{PPTT} 
  E.~Asakawa, S.~Y.~Choi, K.~Hagiwara and J.S. Lee,
  Phys.\ Rev.\ D {\bf 62} (2000) 115005; 
  R.~M.~Godbole, S.~D.~Rindani and R.~K.~Singh,
  Phys.\ Rev.\ D {\bf 67} (2003) 095009; 
  E.~Asakawa and K.~Hagiwara,
  Eur.\ Phys.\ J.\ C {\bf 31} (2003) 351.

\bibitem{GKS}
R.~M.~Godbole, S.~Kraml and R.~K.~Singh,
arXiv:hep-ph/0409199;
arXiv:hep-ph/0501027.


\bibitem{CKLZ}
S.~Y.~Choi, J.~Kalinowski, Y.~Liao and P.~M.~Zerwas,
hep-ph/0407347.


\bibitem{GAbia}
  P.~Garcia-Abia, W.~Lohmann and A.~Raspereza,
  arXiv:hep-ex/0505096;
%
  M.~T.~Dova, P.~Garcia-Abia and W.~Lohmann,
  arXiv:hep-ph/0302113;
%
  P.~Garcia-Abia, W.~Lohmann and A.~Raspereza,
LC-PHSM-2000-062
{\it Prepared for 5th International Linear Collider Workshop (LCWS 
2000),
Fermilab, Batavia, Illinois, 24-28 Oct 2000};
%
  P.~Garcia-Abia and W.~Lohmann,
  Eur.\ Phys.\ J. C {\bf 2} (2000) 2.


\bibitem{BBDS}
  E.~Boos, V.~Bunichev, A.~Djouadi and H.~J.~Schreiber,
  arXiv:hep-ph/0412194.

\bibitem{HagiwaraStong}
  K.~Hagiwara and M.~L.~Stong,
  Z.\ Phys.\ C {\bf 62} (1994) 99.

\bibitem{HIKK}
  K.~Hagiwara, S.~Ishihara, J.~Kamoshita and B.~A.~Kniehl,
  Eur.\ Phys.\ J.\ C {\bf 14} (2000) 457.

\bibitem{CPsuperH}
  J.~S.~Lee, A.~Pilaftsis, M.~Carena, S.~Y.~Choi, M.~Drees, J.~R.~Ellis and
  C.~E.~M.~Wagner, Comput.\ Phys.\ Commun.\  {\bf 156} (2004) 283
  [arXiv:hep-ph/0307377].

\bibitem{TAUPOL}
  B.~Grzadkowski and J.~F.~Gunion,
  Phys.\ Lett.\ B {\bf 350} (1995) 218;
%
  K.~Desch, A.~Imhof, Z.~Was and M.~Worek,
  Phys.\ Lett.\ B {\bf 579} (2004) 157;
%
  A.~Rouge,
  arXiv:hep-ex/0505014.

\bibitem{KL} I.B. Khriplovich and S.K. Lamoreaux, {\em CP Violation
  Without Strangeness} (Springer, New York, 1997).

\bibitem{PR} For a recent review, see, M.~Pospelov and A.~Ritz,
  Annals Phys.\  {\bf 318} (2005) 119.

\bibitem{AR} We thank Adam Ritz for a discussion on this point.

\bibitem{THEDMEXP}
  B.~C.~Regan, E.~D.~Commins, C.~J.~Schmidt and D.~DeMille,
  Phys.\ Rev.\ Lett.\  {\bf 88} (2002) 071805.

\bibitem{APEDM} A. Pilaftsis in Ref.~\cite{CKP}.

\bibitem{APEDMrest} A. Pilaftsis, Phys.\ Lett.\  B {\bf  471} (1999) 
174;
  D. Chang, W.-F. Chang and W.-Y. Keung, Phys.\ Lett.\ B {\bf 478}
  (2000) 239; A. Pilaftsis, Phys.\ Rev.\ D {\bf 62} (2000) 016007. 

\bibitem{FCNC} C.S.  Huang and Q.-S. Yan, Phys.\ Lett.\ B {\bf 442}
 (1998) 209; S.R.~Choudhury and N.~Gaur, Phys.\ Lett.\ B {\bf 451}
 (1999) 86; C.S.  Huang, W.  Liao and Q.-S.  Yan, Phys.\ Rev.\ D {\bf
 59} (1999) 011701; C.~Hamzaoui, M.~Pospelov and M.~Toharia, Phys.\
 Rev.\ D {\bf 59} (1999) 095005; K.S. Babu and C. Kolda, Phys.\ Rev.\
 Lett.\ {\bf 84} (2000) 228.

\bibitem{TEVATRONmumu} 
  D.~Acosta {\it et al.}  [CDF Collaboration],
  Phys.\ Rev.\ Lett.\  {\bf 93} (2004) 032001;
%
  V.~M.~Abazov {\it et al.}  [D0 Collaboration],
  Phys.\ Rev.\ Lett.\  {\bf 94} (2005) 071802.


\bibitem{EOS}
J.~R.~Ellis, K.~A.~Olive and V.~C.~Spanos, arXiv:hep-ph/0504196.

\bibitem{CompHEP}
  A.~Pukhov {\it et al.},
  arXiv:hep-ph/9908288.

\bibitem{PDG}
  S.~Eidelman {\it et al.}  [Particle Data Group],
  Phys.\ Lett.\ B {\bf 592} (2004) 1.



\end{thebibliography}
\end{document}